\documentclass[preprint,12pt]{aastex}

\usepackage{graphicx}
\usepackage{amssymb}
\usepackage{amsmath}
\usepackage{times}
\usepackage[]{eufrak}
\usepackage{color}

\begin{document}

\shorttitle{Effect of Gravitomagnetic Charge on Pulsar
Magnetospheric Structure}

\shortauthors{Morozova, Ahmedov $\&$ Kagramanova}


\title{General Relativistic Effect of Gravitomagnetic Charge on Pulsar
Magnetosphere and Particle Acceleration in a Polar Cap}


\author{V.S. Morozova\altaffilmark{1,2}, B.J. Ahmedov\altaffilmark{1,2}}
\affil{\altaffilmark{1}Institute of Nuclear Physics and Ulugh Begh
Astronomical
Institute, Astronomicheskaya 33, Tashkent 100052, Uzbekistan\\
\altaffilmark{2}International Centre for Theoretical Physics,
Strada Costiera 11, 34014 Trieste, Italy}

\and

\author{V.G. Kagramanova\altaffilmark{3}}
\affil{\altaffilmark{3}Institut f$\ddot{u}$r Physik,
Universit$\ddot{a}$t Oldenburg, D-26111 Oldenburg, Germany}

\altaffiltext{1}{ICTP-OEA Student Fellow}
 \altaffiltext{2}{ICTP
Regular Associate} \altaffiltext{3}{DAAD PhD Student Fellow, on
leave from Ulugh Begh Astronomical Institute, Tashkent 100052,
Uzbekistan}


\begin{abstract}

We study magnetospheric structure surrounding rotating magnetized
neutron star with nonvanishing NUT (Newman-Tamburino-Unti)
parameter. For the simplicity of calculations Goldreich-Julian
charge density is analyzed for the aligned neutron star with zero
inclination between magnetic field, gravitomagnetic field and
rotation axis. From the system of Maxwell equations in spacetime
of slowly rotating NUT star, second-order differential equation
for electrostatic potential is derived. Analytical solution of
this equation indicates the general relativistic modification of
an accelerating electric field and charge density along the open
field lines by the gravitomagnetic charge. The implication of this
effect to the magnetospheric energy loss problem is underlined. In
the second part of the paper we derive the equations of motion of
test particles in magnetosphere of slowly rotating NUT star. Then
we analyze particle motion in the polar cap and show that NUT
parameter can significantly change conditions for particle
acceleration.

\end{abstract}


\section{Introduction}

The theoretical study of radio pulsars can be traced back to the
work of Goldreich $\&$ Julian (1969) where the existence of
magnetosphere with charge-separated plasma around rotating
magnetized neutron stars has been suggested. A spinning magnetized
neutron star generates huge potential differences between
different parts of its surface. The cascade generation of
electron-positron plasma in the polar cap region (Sturrock 1971;
Ruderman $\&$ Sutherland 1975) means that the magnetosphere of a
neutron star is filled with plasma, screening the longitudinal
electric field. This screening results in the corotation of plasma
with a neutron star. Such a rotation is not possible outside the
light cylinder, thus it forms essentially different groups of
field lines: closed, i.e., those returning the stellar surface,
and open, i.e., those crossing the light cylinder and going to
infinity. As a result, plasma may leave the neutron star along the
open field lines and it is generally thought that pulsar radio
emission is produced in the open field line region well inside the
light cylinder (the radius at which the corotation speed equals
$c$).

Study of plasma modes along the field lines is boosted by the
pioneering works of Goldreich $\&$ Julian (1969), Sturrock (1971),
Mestel (1971), Ruderman $\&$ Sutherland (1975), and Arons $\&$
Scharlemann (1979). The subsequent achievements and some new ideas
are reviewed by numerous authors, for example, by Arons (1991),
Michel (1991), Mestel (1992), and Muslimov $\&$ Harding (1997).
Although a self-consistent pulsar magnetosphere theory is yet to
be developed, the analysis of plasma modes in the pulsar
magnetosphere based on the above-mentioned papers provides firm
grounds for the construction of such a model.

The existence of strong electromagnetic fields is one of the most
important features of rotating neutron stars observed as pulsars.
It was shown starting pioneering paper of~\citet{d55} that the
electric field is induced due to the rotation of highly magnetized
star. The general relativistic effect of dragging of inertial
frames is very important in pulsar magnetosphere~\cite{bes90,mt90}
and a source of additional electric field of general relativistic
origin.

The field-aligned electric field is driven by deviation of space
charge from the Goldreich-Julian charge density, which is
determined by the magnetic field geometry. Therefore as it has
been noted by several authors (e.g.~\citet{bes90}, \citet{mt90},
Muslimov \& Harding (1997); Dyks et al (2001), Mofiz $\&$ Ahmedov
(2000)) the general relativistic frame dragging effects on the
field geometry in plasma magnetosphere of rotating neutron stars
is a first-order effect, which has to be carefully included in a
self-consistent model of pulsar magnetospheric structure and
associated electromagnetic radiation. It was first shown by
\citet{mt90} and independently by \citet{bes90} that general
relativistic effects due to frame dragging are crucial for
formation of the field aligned electric field and particle
acceleration in pulsar magnetosphere.

At present there is no any observational evidence for the
existence of gravitomagnetic monopole, that is of exotic
space-time, called NUT space (Newman, Unti and
Tamburino~\cite{nut63}). Therefore it is interesting to study the
electromagnetic fields and processes in NUT space with the aim to
get new tool for studying new important general relativistic
effects which are associated with nondiagonal components of the
metric tensor and have no Newtonian analogues.

\citet{MusTsy92} initiated the detailed general relativistic
derivation of magnetospheric electromagnetic fields around
rotating magnetized neutron star. In this paper, we attempt to
extend work of \citet{MusHar97} by including NUT parameter in
studying particle acceleration along the open field lines of a
rotating neutron star. In \S~\ref{ms} general relativistic
equations describing the electrodynamics of a rotating NUT star
are formulated. The equations are rewritten in the frame of
reference corotating with the rotating NUT star. A detailed
analysis of Goldreich- Julian charge density is also performed. A
general equation governing electrostatic potential in the
magnetosphere is derived.  As it was shown by several authors, in
particular by \citet{ss03}, that the general relativistic effect
on particle motion in pulsar magnetosphere can not be neglected.
Due to this reason  in \S~\ref{acceleration} we are solving
equations of motion for the charged particles for the region near
the magnetic pole just above the stellar surface. From study of
the particle motion along the field lines it has been shown that
the effect of NUT parameter, coupling with the frame-dragging
effect, plays noticeable role in the particle's dynamics. In
\S~\ref{concl} we conclude our findings and discuss them for
further investigations.

Throughout, we use a space-like signature $(-,+,+,+)$ and a system
of units in which $G = 1 = c$ (However, for those expressions with
an astrophysical application we have written the speed of light
explicitly.). Latin indices run $1,2,3$ and Greek ones from $0$ to
$3$.

\section{Plasma Magnetosphere of Slowly Rotating NUT Magnetized
Star} \label{ms}

As it was shown in the paper~\citet{MusTsy92}, from the system of
Maxwell equations, assuming the magnetic field of a neutron star
to be stationary in the corotating frame, the following Poisson
equation for the scalar potential $\Phi$ can be derived
\begin{equation}
\label{Poiss}
{\mathbf\nabla}\cdot\left(\frac{1}{N}{\mathbf\nabla}\Phi\right)=
-4\pi(\rho-\rho_{GJ})\ ,
\end{equation}
where $N\equiv (1-2M/r)^{1/2}$ is the gravitational lapse
function, $M$ is the total mass of the star, $J$ is the angular
momentum of the star, $\rho-\rho_{GJ}$ is the effective space
charge density being responsible for production of unscreened
parallel electric field, $\rho_{GJ}$ is Goldreich-Julian charge
density to be discussed here.

In a pioneering work, Goldreich $\&$ Julian (1969) have shown that
a strongly magnetized, highly conducting neutron star, rotating
about the magnetic axis, would spontaneously build up a charged
magnetosphere. The essence of the argument is that it imposes a
charge magnetosphere that is subject to enormous unbalanced
electric forces being parallel to the magnetic field {\bf B}.
Goldreich \& Julian (1969) hypothesized that a far better
approximation for the magnetosphere would be obtained by shorting
out the component of electric field {\bf{E}} along {\bf{B}}
charges originating in the star. The magnetospheric charges that
maintain ${\bf{E}}\cdot{\bf{B}}$ are themselves subject to the
${\bf{E}}\times{\bf{B}}$ drift that sets them into corotation with
the star. Here we analyze Goldreich-Julian charge density in
general relativity in spacetime of slowly rotating NUT star. An
expression for the Goldreich-Julian charge density $\rho_{GJ}$ can
be calculated in terms of the vector $g_{i}=-g_{0i}/g_{00}$
through the formula
\begin{equation}
\label{rhoGJdef} \rho_{GJ}=-\frac{1}{4\pi}{\mathbf\nabla}(N{\bf
g}\times{\bf B})\ .
\end{equation}

In the metric of the slowly rotating star with the non-vanishing
NUT-parameter
\begin{equation}
\label{metricNUT}
ds^2=-N^2dt^2+N^{-2}dr^2+r^2\left(d\theta^2+\sin^2\theta d\phi^2
\right) -2\left(4N^2l\sin^2\frac{\theta}{2}+\omega
r^2\sin^2\theta\right) d\phi dt
\end{equation}
vector ${\bf g}$ looks like following
\begin{equation}
\label{g} {\bf
g}=\frac{1}{N^2}\left(({\mathbf\Omega}-{\mathbf\omega})\times{\bf
r}- \frac{4lN^2\sin^2\frac{\theta}{2}}{r^2\sin^2\theta}\hat{\bf
z}\times {\bf r}\right)\ .
\end{equation}
This is a stationary axially symmetric solution of vacuum Einstein
field equations with three parameters. The
metric~(\ref{metricNUT}) is the linear approximation of the
Kerr-Taub-NUT metric (see, for example,~\citet{dt02},
~\citet{Bini03}) in the specific angular momentum $a=J/M$ and the
gravitomagnetic monopole $l$. First term in the right hand side of
equation~(\ref{g}) takes into account the well-known effect of
dragging of inertial frames of reference (the Lense-Thirring
effect) with the angular velocity $\omega=2aM/r^3$. Here $\Omega$
is angular velocity of rotation of star. Obviously when $l=0$, the
metric~(\ref{metricNUT}) reduces to the external Hartle-Thorne one
~(\citet{ht68}). On other hand when $\omega$ vanishes the
space-time reduces to the NUT one ~(\citet{nut63}).

To obtain this equation we remember that in the Cartesian
coordinates $\textbf{r}=(x,y,z)$, $x=r\sin\theta\cos\phi$,
$y=r\sin\theta\sin\phi$, $z=r\cos\theta$ and with the $\hat{\bf
z}=(0,0,1)$ the following relation takes place
(see~\citet{Bini03})
\begin{equation}
(\hat{\bf z}\times\textbf{r})\cdot d\textbf{r}=r^2\sin^2\theta
d\phi\ .
\end{equation}

Inserting (\ref{g}) into the expression (\ref{rhoGJdef}) for the
Goldreich-Julian charge density $\rho_{GJ}$ we obtain
\begin{equation}
\label{rhoGJcont}
\rho_{GJ}=-\frac{1}{4
\pi}{\mathbf\nabla}\left\{\frac{1}{N}\left[1-\frac{\kappa}{
\eta^3}-L\left(1-\frac{\varepsilon}{\eta}\right)
\frac{1}{\eta^2}\frac{4\sin^2\frac{\theta}{2}}{\sin^2\theta}\right]
{\bf u}\times{\bf B} \right\}\ ,
\end{equation}
where we denoted $L\equiv {cl}/{\Omega R^2}$, $R$ is the stellar
radius, ${\bf u}={\mathbf\Omega}\times{\bf r}$, $\eta=r/R$ is the
dimensionless radial coordinate, parameter
$\kappa\equiv\varepsilon\beta$, $\varepsilon=2M/R$ is the
compactness parameter, and $\beta=I/I_0$ is the stellar moment of
inertia in the units of $I_0=MR^2$.

In the assumption of the dipole-like configuration for the stellar
magnetic field the nonvanishing components of magnetic field ${\bf
B}$ measured by ZAMO (zero angular momentum observers) with
four-velocity $u_\alpha$
\begin{equation}
u_\alpha = \left\{-N^2,0,0,0\right\}
\end{equation}
take the form (see~\citet{MusTsy92})
\begin{equation}
\label{B} B^{\hat r}=B_0\frac{f(\eta)}{f(1)}\eta^{-3}\cos\theta\ ,
\ \ B^{\hat\theta} =\frac{1}{2} B_0 N\left[-2\frac{f(\eta)}{f(1)}+
\frac{3}{(1-\varepsilon/\eta)f(1)}\right]\eta^{-3}\sin\theta\ ,
\end{equation}
where
\begin{equation}
\label{f}
f(\eta)=-3\left(\frac{\eta}{\varepsilon}\right)^3\left[\ln
\left(1-\frac{\varepsilon}{\eta}\right)+\frac{\varepsilon}{\eta}
\left(1+\frac{\varepsilon}{2\eta}\right)\right]\ ,
\end{equation}
and $B_0\equiv 2\mu/R^3$ is the Newtonian value of the magnetic
field at the pole of star, hat labels the orthonormal components,
$\mu$ is the magnetic moment. The solution of this type was first
obtained by \citet{go64} and then reproduced by the number of
authors.

The polar angle $\Theta$ of the last open magnetic line as a
function of $\eta$ will look like (see~\citet{mt91}
or~\citet{MusTsy91})
\begin{equation}
\label{Theta}
\Theta\cong\sin^{-1}\left\{\left[\eta\frac{f(1)}{f(\eta)}\right]^{1/2}
\sin\Theta_0\right\}\ ,\
\Theta_0=\sin^{-1}\left(\frac{R}{R_{LC}f(1)}\right)^{1/2}\ .
\end{equation}
Parameter $\Theta_0$ has the meaning of the magnetic colatitude of
the last open magnetic line at the stellar surface,
$R_{LC}=c/\Omega$ is the light-cylinder radius.

Making further algebraic transformations on the equation
(\ref{rhoGJcont}) and taking into account equation (\ref{B}) we
come to the following expression for the Goldreich-Julian charge
density

\begin{equation}
\label{rhoGJ} \rho_{GJ}=-\frac{\Omega B_0}{2\pi c
}\frac{1}{N\eta^3}\frac{f(\eta)}{f(1)}\left\{1-\frac{\kappa}{\eta^3}-
L\left(1-\frac{\varepsilon}{\eta}\right)\frac{1}{\eta^2}
\frac{4\sin^2\frac{\theta}{2}}{\sin^2\theta}\right\}\ .
\end{equation}

\begin{figure}
\plotone{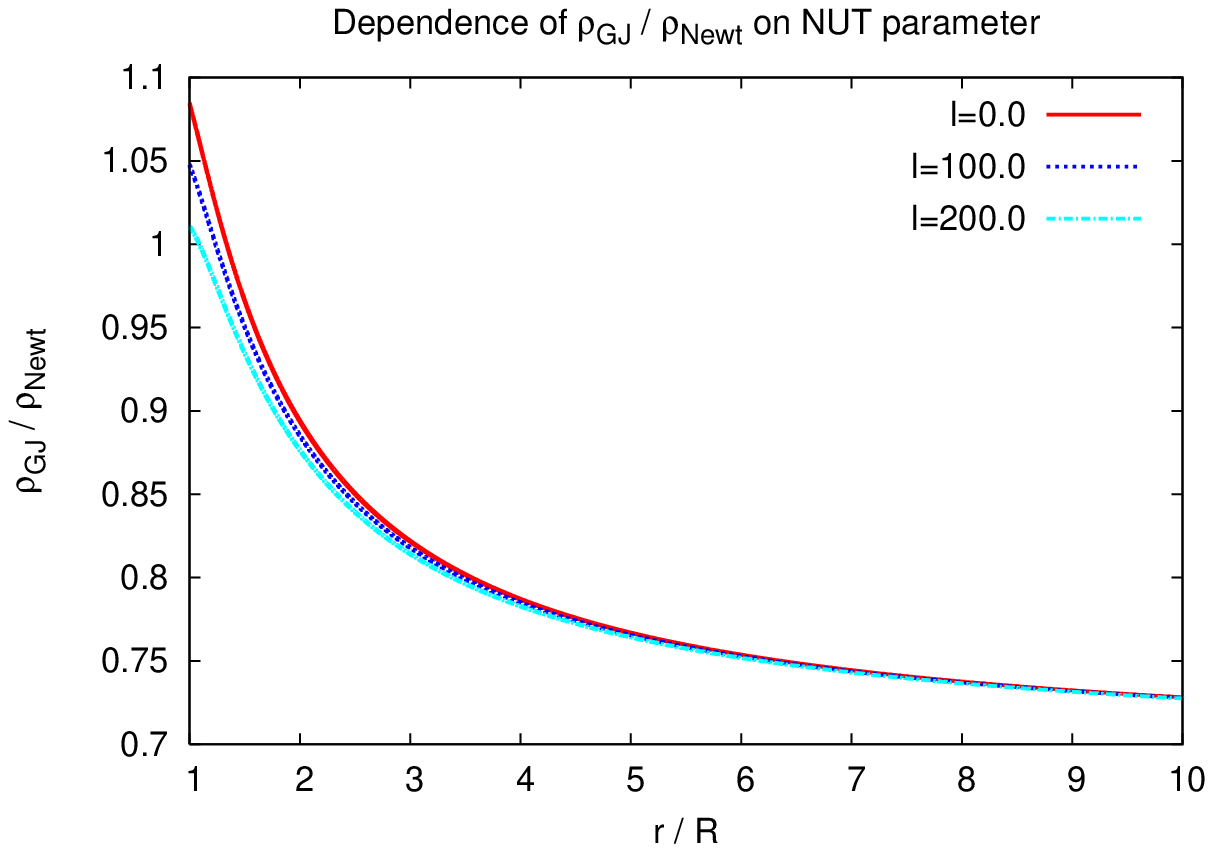} \caption{Radial dependence of the
Goldreich-Julian charge density normed in its Newtonian expression
for the different values of the NUT parameter.} \label{fig1}
\end{figure}

Hereafter for the simplicity of calculations we assume that the
inclination angle between the magnetic axis and the axis of
rotation of the star is equal to zero. Figure~\ref{fig1} presents
the radial dependence of obtained general relativistic expression
for the Goldreich-Julian charge density $\rho_{GJ}$ normed in its
Newtonian expression for several different values of NUT
parameter. It can be found from the figure that even for
comparatively small values of NUT parameter its influence on the
Goldreich-Julian charge density $\rho_{GJ}$~(\ref{rhoGJ}) plays an
important role. The value of the Goldreich-Julian charge density
$\rho_{GJ}$ at the surface of the star is more sensitive to the
NUT parameter. The tendency seen from the Figure is so that
$\rho_{GJ}$ is vise proportional to the NUT parameter at the
surface of the star and in asymptotics it reaches Newtonian
expression. Here we took typical numbers for neutron star
parameters as $R=10 km$, $M=2 km$ and $T=0.1 s$.

For the relativistic plasma the charge density $\rho$ is
proportional to magnetic field with the proportionality
coefficient being constant along the given magnetic field line
(see, for example, \citet{MusTsy91}) that is
\begin{equation}
\label{rho} \rho=\frac{\Omega B_0}{2\pi c
}\frac{1}{N\eta^3}\frac{f(\eta)}{f(1)}A(\xi)\ ,
\end{equation}
where $\xi=\theta/\Theta$ is the dimensionless angular variable,
$A(\xi)$ is an unknown function to be defined from the boundary
conditions. One could  insert expressions (\ref{rhoGJ}) and
(\ref{rho}) into the Poisson equation (\ref{Poiss}), and in the
approximation of small angles $\theta$ get the following
differential equation
\begin{eqnarray}
R^{-2}\left\{N\frac{1}{\eta^2}\frac{\partial}{\partial\eta}\left(\eta^2
\frac{\partial}{\partial\eta}\right)+\frac{1}{N\eta^2\theta}\left[
\frac{\partial}{\partial\theta}\left(\theta
\frac{\partial}{\partial\theta}\right)+\frac{1}{\theta}
\frac{\partial^2}{\partial\phi^2}\right]\right\}\Phi \nonumber
\\
=-4\pi\frac{\Omega B_0}{2\pi
c}\frac{1}{N\eta^3}\frac{f(\eta)}{f(1)}\left\{1-\frac{\kappa}{\eta^3}-
L\left(1-\frac{\varepsilon}{\eta}\right)\frac{1}{\eta^2}+A(\xi)\right\}\
.
\end{eqnarray}

Here we used a fact that in the small-angle limit
\begin{equation}
\label{approx} \frac{4\sin^2\frac{\theta}{2}} {\sin^2\theta}\sim1\
.
\end{equation}

Our further discourses are based on the extension of
work~\citet{MusTsy92} to NUT spacetime. Using dimensionless
function $F=\eta\Phi/\Phi_0$, where $\Phi_0=\Omega B_0 R^2$ and
variables $\eta$ and $\xi$, we can rewrite the equation
(\ref{Poiss}) for the dimensionless electrostatic potential as
\begin{equation}
\label{difLam}
\left[\frac{d^2}{d\eta^2}+\Lambda^2(\eta)\frac{1}{\xi}\frac{\partial}
{\partial\xi}
\left(\xi\frac{\partial}{\partial\xi}\right)\right]F=
-\frac{2}{\eta^2\left(1-\frac{\varepsilon}{\eta}\right)}\frac{f(\eta)}{f(1)}
\left[1-\frac{\kappa}{\eta^3}-
L\left(1-\frac{\varepsilon}{\eta}\right)\frac{1}{\eta^2}+A(\xi)\right]\
,
\end{equation}
where
$\Lambda(\eta)=[\eta\Theta(\eta)(1-\varepsilon/\eta)^{1/2}]^{-1}$.

After performing Fourier-Bessel transformation
\begin{equation}
\label{FourBes}
F(\eta,\xi)=\sum_{i=1}^{\infty}F_i(\eta)J_0(k_i\xi)\ , \quad
F_i(\eta)=\frac{2}{[J_1(k_i)]^2}\int^1_0\xi F(\eta,\xi)J_0
(k_i\xi)d\xi\ ,
\end{equation}
and using relation
\begin{equation}
\sum_{i=1}^{\infty}\frac{2}{k_i J_1(k_i)}J_0(k_i\xi)=1\ ,
\end{equation}
one can obtain equation (\ref{difLam}) in the form
\begin{equation}
\label{difgamma}
\left(\frac{d^2}{d\eta^2}-\gamma^2_i(\eta)\right)F_i=
-\frac{2}{\eta^2\left(1-\frac{\varepsilon}{\eta}\right)}\frac{f(\eta)}{f(1)}
\left[\frac{2}{k_i
J_1(k_i)}\left\{1-\frac{\kappa}{\eta^3}-L\left(1-\frac{\varepsilon}{\eta}\right)
\frac{1}{\eta^2}\right\}+A_i\right]\ ,
\end{equation}
where $\gamma^2_i=k^2_i\Lambda^2$, $k_i$ are positive zeros of the
functions $J_0$.

Considering a region near the surface of the star, where
$z=\eta-1\ll1$, and using following boundary conditions (that is
the conditions of equipotentiality of the stellar surface and zero
steady state electric field at $r=R$)
\begin{equation}
F_i|_{z=0}=0\ ,\ \frac{\partial F_i}{\partial z}|_{z=0}=0\ .
\end{equation}
one can find the expression for the scalar potential $\Phi$ near
the surface of the star and corresponding to this potential
component of the electric field $E_{\|}$, being parallel to the
magnetic field (see the discourses of~\citet{MusTsy92} work):
\begin{equation}
\Phi=\frac{36\Phi_0}{\eta}\sqrt{1-\varepsilon}(\kappa-L\varepsilon)
\Theta^3_0\sum^{\infty}_{i=0}\left[\exp\left\{\frac{k_i(1-\eta)}
{\Theta_0\sqrt{1-\varepsilon}}\right\}-1+\frac{k_i(\eta-1)}
{\Theta_0\sqrt{1-\varepsilon}}
\right]\frac{J_0(k_i\xi)}{k_i^4J_1(k_i)}\ ,
\end{equation}
\begin{equation}
E_{\|}=-\frac{36\Phi_0}{R}(\kappa-L\varepsilon)\Theta^2_0
\sum^{\infty}_{i=0}\left[1-\exp\left\{\frac{k_i(1-\eta)}
{\Theta_0\sqrt{1-\varepsilon}}\right\}\right]
\frac{J_0(k_i\xi)}{k_i^3J_1(k_i)}\ .
\end{equation}

It should be noted that these formulae differ from the ones
obtained in~\citet{MusTsy92} by that the parameter $\kappa$ gets
replaced by $(\kappa-L\varepsilon)$.

Considering now the region $\Theta_0\ll\eta-1\ll R_{LC}/R$, where
$|d^2F_i/d\eta^2|\ll\gamma^2_i(\eta)|F_i|$ one can see that
equation (\ref{difgamma}) becomes
\begin{equation}
-\gamma^2_i(\eta)F_i=
-\frac{2}{\eta^2\left(1-\frac{\varepsilon}{\eta}\right)}\frac{f(\eta)}{f(1)}
\left[\frac{2}{k_i
J_1(k_i)}\left\{1-\frac{\kappa}{\eta^3}-L\left(1-\frac{\varepsilon}{\eta}\right)
\frac{1}{\eta^2}\right\}+A_i\right]\ ,
\end{equation}
from which it immediately follows
\begin{eqnarray}
F_i=\frac{2}{k^2_i}\theta^2(\eta)\frac{f(\eta)}{f(1)}
\left[\frac{2}{k_i J_1(k_i)}\left\{
(\kappa-L\varepsilon)\left(1-\frac{1}{\eta^{3}}-\frac{3}{\gamma_i(1)}\right)
+L\left(1-\frac{1}{\eta^2}\right) \right\}\right]\ .
\end{eqnarray}

Using this expression for $F_i$ one can obtain the scalar
potential in the region at distances greater than the polar cap
size as
\begin{eqnarray}
\Phi&=&\frac{\Phi_0}{\eta}F=2\Phi_0\Theta^2_0(\kappa-L\varepsilon)
\left(1-\frac{1}{\eta^3}\right)\sum_i\frac{2J_0(k_i\xi)}{k^3_i
J_1(k_i)}\nonumber \\
&=&\frac{1}{2}\Phi_0\Theta_0^2(\kappa-L\varepsilon)
\left(1-\frac{1}{\eta^3}\right)(1-\xi^2)\nonumber \\
&=&\frac{1}{2} \Omega R^2 B_0\Theta_0^2(\kappa-L\varepsilon)
\left(1-\frac{1}{\eta^3}\right)(1-\xi^2)\ .
\end{eqnarray}

\begin{figure}
\plotone{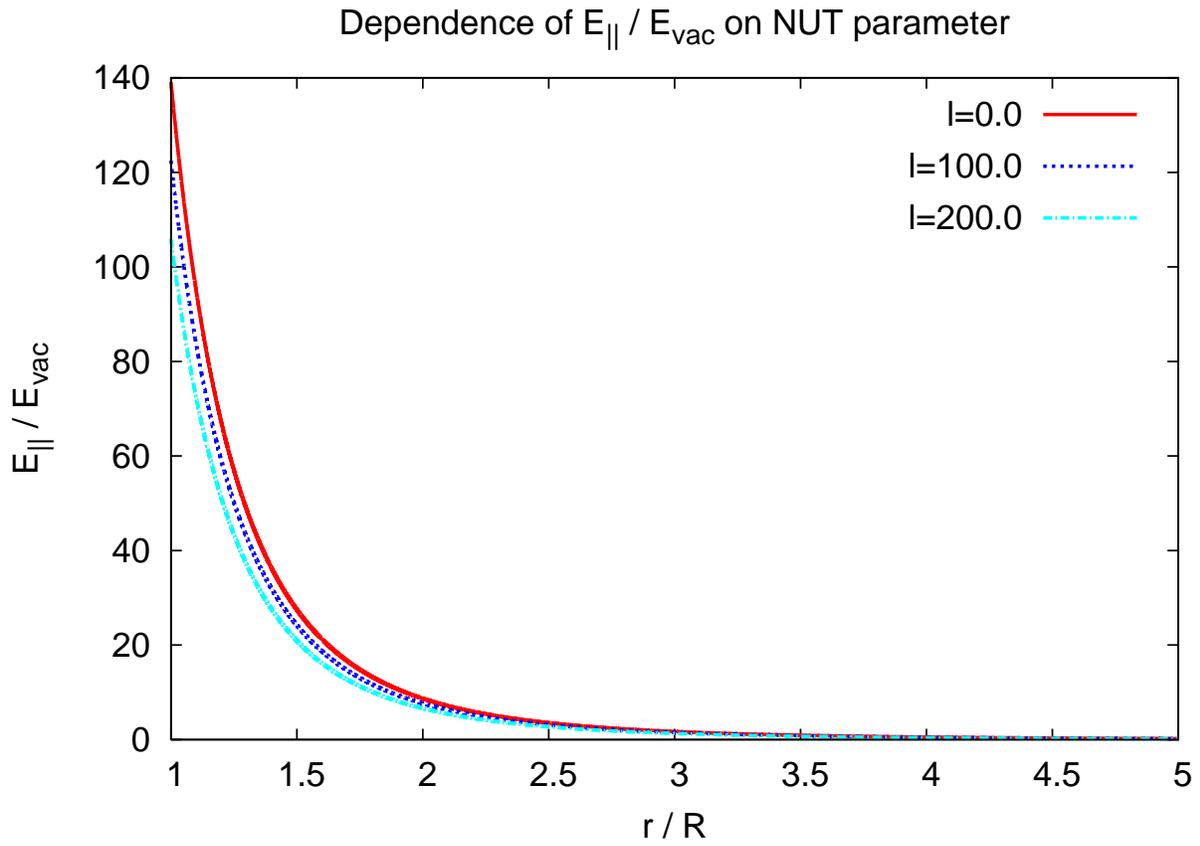} \caption{Radial dependence of the accelerating
component of electric field normed in its vacuum magnitude for the
different values of the NUT parameter.} \label{fig2}
\end{figure}

Corresponding to this potential component of electric field
$E_{\|}$ will look like
\begin{equation}
E_{\|}=-\frac{1}{R}\frac{\partial\Phi}{\partial\eta}|_{\xi=constant}=-E_{vac}
\Theta^2_0\frac{3(\kappa-L\varepsilon)}{2\eta^4}(1-\xi^2)\ ,
\end{equation}
where $E_{vac}\equiv(\Omega R/c)B_0$ is the characteristic
Newtonian value of the electric field generated near the surface
of a neutron star rotating in vacuum~\citet{d55}. In
Fig.~\ref{fig2} one can find the radial dependence of parallel
electric field $E_{\|}$ in terms of $E_{vac}$ for the different
values of NUT parameter.

The energy losses from the polar cap of the rotating star with
non-vanishing NUT-parameter can now be calculated. According
to~\citet{MusHar97} the expression for the total power carried
away by relativistically moving particles is
\begin{equation}
\label{Ldif} L_p=2(-c\int\rho\Phi\ dS)\ .
\end{equation}

In slow rotating NUT spacetime
\begin{equation}
\label{rphi}
-\rho\Phi\approx\frac{1}{4\pi}\left(\frac{\Omega
B_0}{c}\right)^2\frac{R^2\Theta^2_0}{N\eta^3}\frac{f(\eta)}{f(1)}
\left[(\kappa-L\varepsilon)\left(1-\kappa-
L(1-\varepsilon)\right)\right](1-\xi^2)\ .
\end{equation}

Inserting (\ref{rphi}) into (\ref{Ldif}) and taking the integral
one can get, that
\begin{equation}
\label{Lmax}
(L_p)_{max}=\frac{3}{2}\left[(\kappa-L\varepsilon)
\left(1-\kappa-L(1-\varepsilon)\right)\right]\dot{E}_{rot}\ ,
\end{equation}
where
\begin{equation}
\dot{E}_{rot}\equiv\frac{1}{6}\frac{\Omega^4 B_0^2 R^6}{c^3
f^2(1)}=\frac{1}{f^2(1)}(\dot{E}_{rot})_{Newt}\
\end{equation}
and $(\dot{E}_{rot})_{Newt}$ is the standard Newtonian expression
for the magneto-dipole losses in flat space-time approximation.

\begin{figure}
\plotone{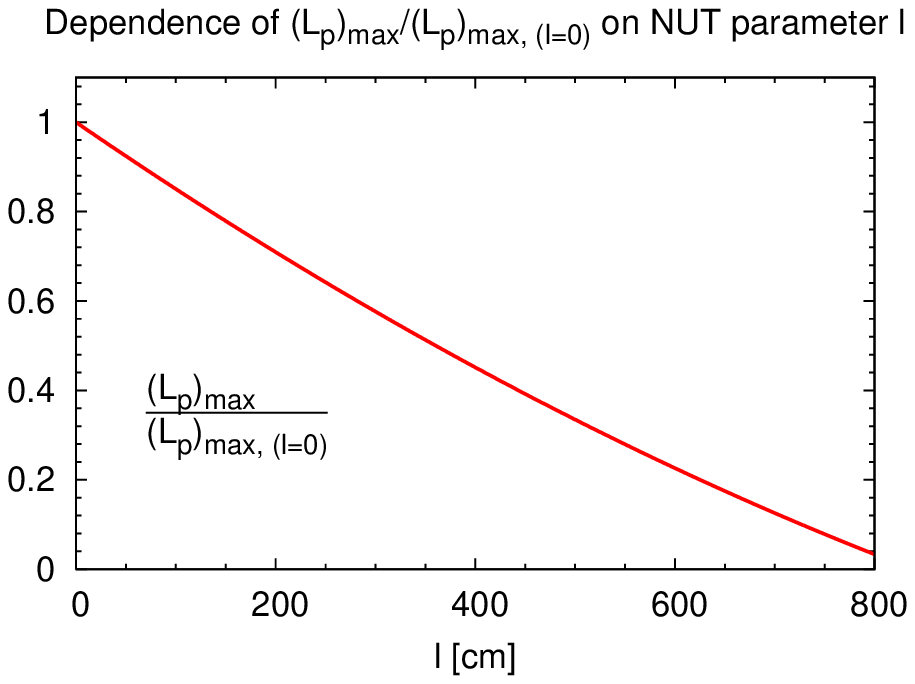} \caption{Ratio of polar cap energy losses into
its zero gravitomagnetic charge expression as a function of NUT
parameter.}  \label{fig3}
\end{figure}

In the limiting case when $l\rightarrow 0$ one could get the
result~\citet{MusHar97}:
\begin{equation}
\label{Lmax0} (L_p)_{max\
(l=0)}=\frac{3}{2}\kappa(1-\kappa)\dot{E}_{rot}\ .
\end{equation}
The ratio
\begin{equation}
\label{ratioL} \frac{(L_p)_{max}}{(L_p)_{max\
(l=0)}}=1-\frac{L(\kappa+\varepsilon
-2\kappa\varepsilon)}{\kappa(1-\kappa)}+\frac{L^2\varepsilon(1-\varepsilon)}
{\kappa(1-\kappa)}\ .
\end{equation}
as function of NUT parameter is presented in Fig.~\ref{fig3}. {The
dependence has
 a parabolic form. Namely, for small values of NUT parameter the energy
  losses are decreasing. Then the graph has a minimum and with further growth of NUT parameter,
  more energy is lost.} Physically it is due to the fact that
  contributions in accelerating electric field coming from the
  different parameters have the opposite signs.

Taking into account that $\kappa=\varepsilon\beta\sim\varepsilon$
one can see, that the magnitude of additional terms arising from
the NUT parameter is determined by the magnitude of $L$. For a
millisecond pulsar with $l\sim10^3cm$, $\Omega\sim10s^{-1}$ and
$R\sim10^6cm$ we have
\begin{equation}
L=\frac{c l}{\Omega R^2}\sim 1\ .
\end{equation}
It shows that corrections to energy losses, concerned with
non-zero NUT parameter cannot be neglected and in principle can
provide an important information which will help to detect the
gravitomagnetic charge.

{Equation (\ref{Lmax}) has physical sense only if $(L_p)_{max} <
\dot{E}_{rot}$, that means the power of a polar cap accelerator in
principle cannot exceed the total spin-down power of a pulsar if
one suppose that electromagnetic radiation is powered by rotation.
Using equations (\ref{Lmax}), (\ref{Lmax0}) and (\ref{ratioL}) one
can see that the physical sense is preserved for discussed
magnetosphere model (when electric field created by rotation of
star is dominating) if
\begin{equation}
L^2\lesssim\frac{2}{3\kappa(1-\kappa)}-1\ .
\end{equation}
Using the value $\kappa=0.15$ (\cite{MusTsy92}) one can obtain
upper limit for the value of NUT-parameter $l\lesssim ~1000 cm$.}

For practical useful applications the equation (\ref{Lmax}) can be
rewritten in terms of pulsar's observable characteristics as the
period $P$ and its time derivative $\dot{P}\equiv dP/dt$:
\begin{equation}
\label{PP} (P\dot{P})_{max}=\frac{3}{4}\left[(\kappa-L\varepsilon)
\left(1-\kappa-L(1-\varepsilon)\right)\right]\frac{I}{\tilde{I}}
\frac{1}{f^2(1)}(P\dot{P})_{Newt}\ ,
\end{equation}
where the expressions
\begin{equation}
\label{LPP} (L_p)_{max}=-\tilde{I}(\Omega\dot{\Omega})_{max}
\end{equation}
and
\begin{equation}
(P\dot{P})_{Newt}\equiv\left(\frac{2\pi^2}{3c^3}\right)
\frac{R^6B^2_0}{I}\
\end{equation}
have been taken into account.

In equation (\ref{LPP}) $\tilde{I}$ is the general relativistic
moment of inertia of the star (see e.g.~\cite{r3})
\begin{equation}
\tilde{I}\equiv\int d^3\mathbf{x}\sqrt{\gamma}e^{-\Phi(r)}\rho
r^2\sin^2\theta\ ,
\end{equation}
where $e^{-\Phi(r)}\equiv 1/\sqrt{-g_{00}}$, $\rho(r)$ is the
total energy density, $\gamma$ is the determinant of the three
metric and $d^3\mathbf{x}$ is the coordinate volume element.

Period of pulsar and it's time derivative are very precisely
measured quantities for a large number of pulsars (for example, in
the paper of~\cite{Kaspi04} there is a $P-\dot{P}$ diagram for the
1403 catalogued rotation-powered pulsars, see also~\cite{a07}).
Thus, expression (\ref{PP}) for $P\dot{P}$ may indicate the
possible existence and magnitude of NUT-parameter. The main
difficulty encountered on this way nowadays is the uncertainty of
estimation of the moment of inertia of the neutron star. But in
future, when the moment of inertia of the neuron stars would be
determined more precisely, one can in principle obtain the value
of NUT-parameter from the observational data.

\section{Charged Particle Acceleration in a Polar Cap in Magnetosphere
of Slowly Rotating NUT Star} \label{acceleration}

The origin of radio emission from the polar cap is one of the most
mysterious, still unsolved question in the physics of pulsars. One
of the likely scenarios is that particles are accelerated along
open magnetic field lines and emit $\gamma$-rays that subsequently
convert into electron-positron pairs under a strong magnetic
field. The combination of the primary beam and pair plasma
provides the radio emission mechanism. Due to this reason it is
interesting to study particle acceleration conditions and
equations of motion in the pulsar magnetosphere in the presence of
NUT parameter.

In our preceding paper~\citet{ma00} effects of general relativity
on plasma modes along the open field lines of rotating magnetized
neutron star have been studied. Here we investigate equations of
motion of charged particle in the region just above the polar cap
surface of a neutron star with NUT parameter by extending results
of~\citet{ss03} to spacetime with gravitomagnetic charge.

Equations of motion for particle with mass $m$, charge $e$, proper
time $\tau$ and 4-velocity $v^{\mu}\equiv dx^{\mu}/d\tau$ look
like
\begin{equation}
\label{motion}
m\left(\frac{dv^{\mu}}{d\tau}+\Gamma^{\mu}_{\nu\lambda}v^{\nu}v^{\lambda}\right)
=e \mathcal{F}^{\mu\nu}v_{\nu} \ ,
\end{equation}
where $\Gamma^{\mu}_{\nu\lambda}$ is the affine connection,
$\mathcal{F}^{\mu\nu}=u^\mu E^\nu -u^\nu E^\mu
+\eta^{\mu\nu\gamma\rho}u_\gamma B_\rho$ is the electromagnetic
field tensor, $\eta^{\mu\nu\gamma\rho}$ is the Levi-Civita tensor.

 In the approximation of small angles
$\theta$ in the polar cap region the equations of
motion~(\ref{motion}) can be rewritten in the form
\begin{equation}
\label{eqmot1}
 \frac{N}{s^2}\frac{d}{d s}\left(s^2\frac{d\phi}{d
s}\right)-\frac{l(l+1)}{N s^2}\phi=\frac{B}{B_0
N}\left(\frac{j}{V}-\bar{j}\right)\ ,
\end{equation}
\begin{equation}
\label{eqmot2} \frac{d}{d s}\left(N\gamma\right)=\frac{1}{N^2}
\frac{d\phi}{d s}\
\end{equation}
in the space-time outside the star described by Kerr-Taub-NUT
metric in the slow rotation limit (\ref{metricNUT}). It formally
coincides with the equations of motion from~\citet{ss03}. Here the
Lorentz factor $\gamma\equiv -v^\mu u_\mu =Nv^0=1/(1-V^2)^{1/2}$,
the relative 3-velocity $V^i\equiv v^\mu h^i_\mu/\gamma$, the
projection tensor $h^{\alpha\beta}\equiv g^{\alpha\beta}+ u^\alpha
u^\beta$ and the following normalized variables
%
\begin{eqnarray}
\label{var}
 j\equiv-\frac{2\pi N(s) J(s)}{\Omega B(s)}=const\ ,\quad
\phi(s)\equiv\frac{e}{m}\Phi(s)\ ,\nonumber \\
s\equiv\sqrt{\frac{2\Omega B_0e}{mc^2}}r\ ,
\quad\bar{j}\equiv-\frac{2\pi N(s)\rho_{GJ}(s)}{\Omega B(s)}\
\end{eqnarray}
are introduced. To obtain equations (\ref{eqmot1}) and
(\ref{eqmot2}) the scalar potential $\Phi$ was expanded in spherical
harmonics $Y_{lm}(\theta,\phi)$ as
$\Phi=\sum_{l,m}\bar{\Phi}(r)Y_{lm}(\theta,\phi)$ and only the mode
of the polar cap scale $l\approx\pi/\Theta_0$ was taken. In
(\ref{var}) $J$ is the electric charge current density,
$B\equiv|\textbf{B}|=B^{\hat r}+O(\theta^2)$.

The approximate value of Goldreich-Julian charge density
(\ref{rhoGJcont}) is
\begin{equation}
\label{rGJapp}
\rho_{GJ}=\frac{1}{2\pi}\frac{B\Omega}{N}\left(1-\frac{\omega}{\Omega}-\frac
{N^2l}{\Omega}\right)\ ,
\end{equation}
and consequently NUT-parameter will provide additional radial
dependence to $\bar{j}$.

According to recent paper of~\cite{Bel07} motion of charged
particles in the vicinity os the polar cap of the pulsar is
governed by equation (momentum of the particle is given in units
of $mc$)
\begin{equation}
\frac{dp}{dt}=\frac{eE_\|}{mc}\ .
\end{equation}

And for parallel electric field $E_\|$ one could get equation
\begin{equation}
\label{Gauss} \nabla\cdot
E=\frac{dE_\|}{dz}=4\pi(\rho-\rho_{GJ})\,\qquad z\ll r_{pc}\ ,
\end{equation}
where $r_{pc}$ is the radii of the polar cap of the star.
Rewriting (\ref{Gauss}) using $d/dt\equiv vd/dz$, $a\equiv
j/c\rho_{GJ}$, $\rho=j/v$ and $v=cp(1+p^2)^{-1/2}$ one could
obtain (see ~\cite{Bel07})
\begin{equation}
\frac{dE_\|}{dt}=4\pi j\left(1-\frac{ap}{\sqrt{1+p^2}}\right)\ ,
\end{equation}
where
\begin{equation}
\label{a} a(z)=
a_0\frac{1-\kappa+L\varepsilon}{(1-\kappa+L\varepsilon)(1+z/R)^{-3}}\
.
\end{equation}

\begin{figure}
\plotone{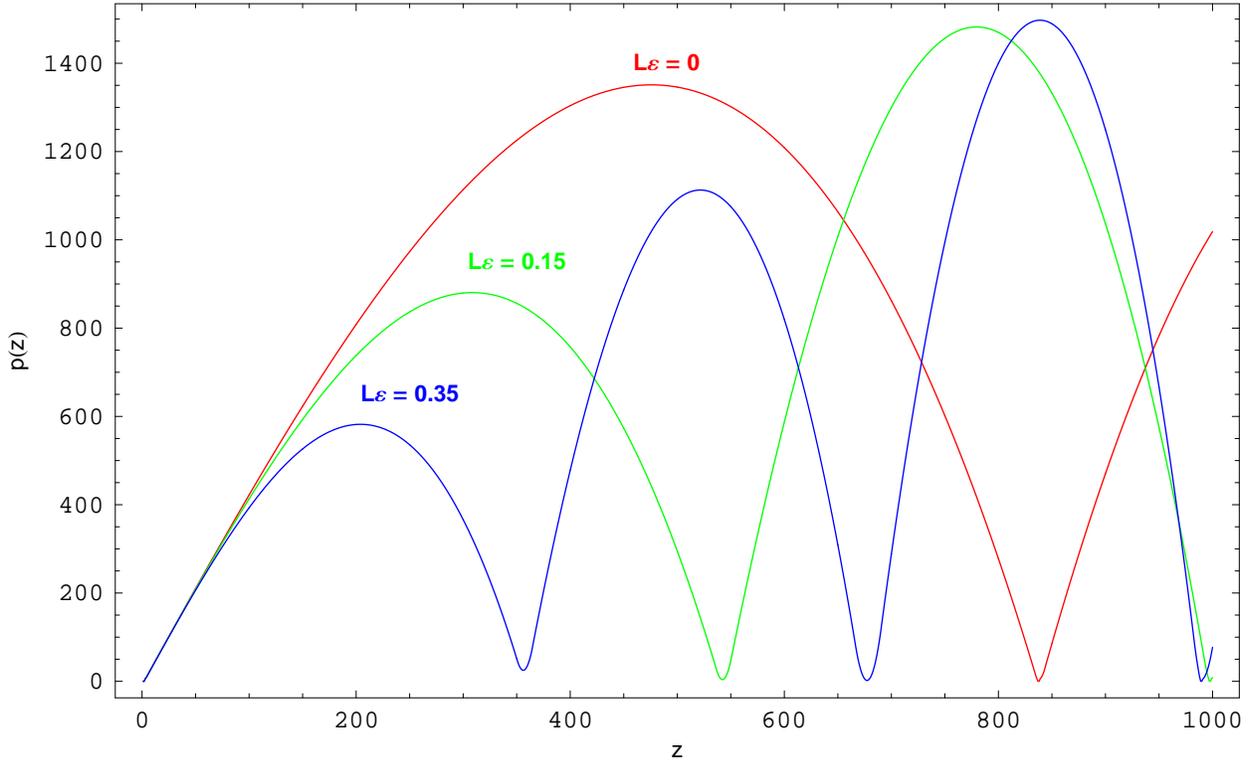} \caption{The dependence of the momentum of
charged particle, extracted from the polar cap by $E_\|$, on
height for different values of NUT-parameter. The range of used
values: $B=3\times10^{12}G$, $P=1s$, $k=0.15$, $a_0=0.999$,
rotator is assumed to be aligned. Considered region is $z<100m$
which lies within the polar cap radii.} \label{fig 0999}
\end{figure}

\begin{figure}
\plotone{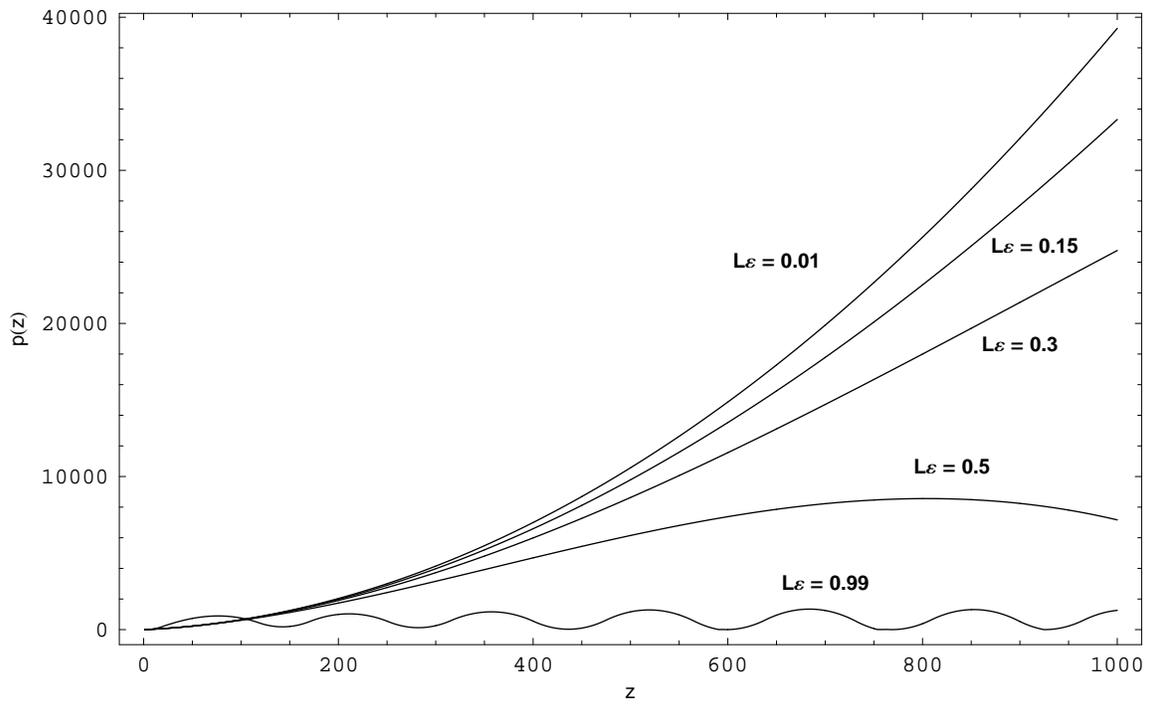} \caption{The dependence of the momentum of
charged particle, extracted from the polar cap by $E_\|$, on
height for $a_0=1.1$, $k=0$ and different values of
NUT-parameter.} \label{fig 11}
\end{figure}

\begin{figure}
\plotone{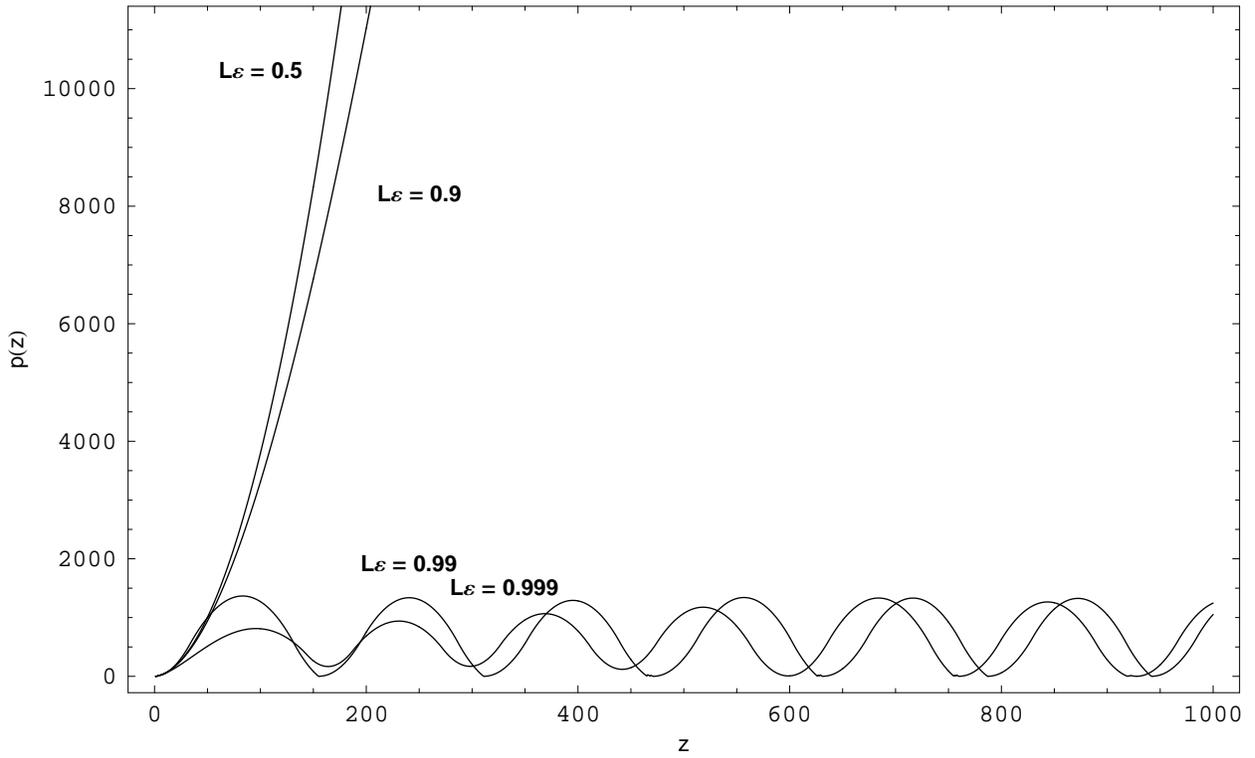} \caption{The dependence of the momentum of
charged particle, extracted from the polar cap by $E_\|$, on
height for $a_0=2$, $k=0$ and different values of NUT-parameter.
When NUT-parameter $L\varepsilon$ is very close to $1$,
oscillations take place independently on the value of $a_0$.}
\label{fig 2}
\end{figure}

Then the system of equations for $p(z)$ is graphically solved.
Fig.\ref{fig 0999} presents $p(z)$ for several values of
$L\varepsilon$, when $\kappa=0.15$ (\cite{MusTsy92}) and
$a_0=0.999$ and shows that influence of NUT-parameter
significantly changes the period of oscillations. From the
Fig.~\ref{fig 11} and Fig.~\ref{fig 2} one could see when
$\kappa-L\varepsilon\sim 1$ oscillations take place even for large
$a_0$. For the values of $\kappa-L\varepsilon$ very close to $1$,
the character of graphs almost has no dependence on the value of
$a_0$. For old recycled neutron stars with almost zero angular
momentum the effects connected with NUT-parameter may play single
mechanism for pulsar radiation.

Following~\cite{a98} and using equations (\ref{rhoGJ}),
(\ref{rho}) and (\ref{approx}),  one can obtain for the
accelerating potential drop:
\begin{equation}
\Delta\Phi_{\|}\approx(\kappa-L\varepsilon)\Phi_{pole}
\left[1-\left(\frac{1}{\eta^3}\right)\right]\ ,
\end{equation}
while in Newtonian case it looks like
\begin{equation}
\Delta\Phi_{\|}\approx\Phi_{pole}\left(\frac{R}{\rho_B}\right)\ ,
\end{equation}
where $\rho_B$ is the radius of curvature of the magnetic field
lines, $\Phi_{pole}\equiv\Omega^2\mu/c^2$ and $\mu$ is the
magnetic moment.

Taking into account the frame dragging effect noticeably improve
comparison between theory and observations. As
$R/\rho_B\sim10^{-2}P^{-1/2}$ and $\kappa\sim10 P^{-1/2}$, the
energy of curvature gamma rays in relativistic frame-dragging case
may rise up to be $1000$ times greater than that occurs
in~\cite{as79} pair creation theory (see ~\cite{a98}). As the
effect of NUT-parameter seems to have the same order as the effect
of frame dragging, it makes an additional contribution to the
energy of curvature gamma photons and thus may play noticeable
role in the formation of the plasma magnetosphere of neutron star
with gravitomagnetic charge.

\section{Conclusion}
\label{concl}

We have considered astrophysical processes in the polar cap of
pulsar magnetosphere in assumption of slowly rotating NUT
space-time. In particular, general-relativistic corrections to the
Goldreich-Julian charge density, electrostatic scalar potential
and accelerating component of electric field being parallel to
magnetic field lines in the polar cap region due to the presence
of gravitomagnetic charge are found. The presence of NUT-parameter
slightly modulates Goldreich-Julian charge density near the
surface of the star. However as it is known the effective electric
charge density i.e.  difference between Goldreich-Julian charge
density $\rho_{GJ}$  (being proportional in the case of flat
space-time to $\mathbf{\Omega\cdot B}$) and electric charge
density (being proportional to $\mathbf{B}$) in the star
magnetosphere is responsible for the generation of electric field
being parallel to magnetic field lines. This difference is equal
to zero at the surface of the star and changes with the distance
from it due to the fact that $\rho$ can not compensate
$\rho_{GJ}$. General relativitistic terms arising from the
dragging of inertial frames and presence of the gravitomagnetic
charge give very important additional contribution to this
difference. Both of these terms depend on the radial distance from
the star as $1/r^3$ and have equally important influence on the
value of accelerating electric field component generated in the
magnetosphere near the surface of the neutron star.

These results are applied to find an expression for
electromagnetic energy losses along the open magnetic field lines
of the slowly rotating NUT star. It is found that in the case of
non-vanishing NUT-parameter an additional important term to the
coefficient of standard magneto-dipole energy losses expression
appears. Comparison of effect of NUT-parameter with the already
known effects shows that it can not be neglected. Obtained new
dependence may be combined with astrophysical data on pulsar
periods slow down and be useful in further investigations on
possible detection of the gravitomagnetic monopole.

It is also shown that the presence of gravitomagnetic charge has
influence on the conditions of particles motion in the polar cap
region. From derived results it can be seen, that NUT-parameter
modulates the period of oscillations of particle's momentum. As
the effect, connected with NUT-parameter has the angular velocity
of the star's rotation in denominator, it will be increased for
the old stars, where the frame-dragging effect will be decreased.
Thus possible candidates for verification of the existence of NUT
parameter are old compact objects with extremely low rotation
period.

\acknowledgments

This research is supported in part by the UzFFR (project 05-08)
and projects FA-F2-F079, FA-F2-F061 and A13-226 of the UzAS, by
the ICTP through the OEA-PRJ-29 and by NATO through the
reintegration grant EAP.RIG.981259. VGK acknowledges the financial
support from DAAD. BJA is grateful to Exchange of Astronomers
Programme of the IAU C46-PG-EA and the ICTP Associate Programme
for the travel support. Authors thank Ahmadjon Abdujabbarov for
his help and assistance in making numerical calculations and
graphs and anonymous referee for his invaluable work in improving
our manuscript.

\clearpage

\newpage

\newpage

\newpage
\newpage

\newpage

%



\label{lastpage}

\clearpage


\begin{thebibliography}{Newman, Tamburino \& Unti(1963)}

\bibitem[Arons (1998)]{a98}
Arons, J. 1998, arXiv:astro-ph/9802198 v3

\bibitem[Arons(2007)]{a07}
Arons, J. 2007, arXiv:0708.1050v1 [astro-ph]

\bibitem[Arons \& Scharlemann(1979)]{as79}
Arons, J., \& Scharlemann, E. T. 1979, ApJ, 231, 854

\bibitem[Beloborodov(2007)]{Bel07}
Beloborodov, A. M. 2007, arXiv:0710.0920v1 [astro-ph]

\bibitem[Beskin(1990)]{bes90}
Beskin, V. S. 1990, Soviet Astron. Lett., 16, 286

\bibitem[Beskin(2005)]{bes05}
Beskin, V. S. 2005,  "Osesimmetrichnye  Stacionarnye Techeniya v
Astrofizike"   (Axisymmetric Stationary Flows in Astrophysics),
Moscow, Fizmatlit, in Russian

\bibitem[Bini et al.(2003)]{Bini03}
Bini, D., Cherubini, C., Jantzen, R. T., \& Mashhoon, B. 2003,
Class. Quantum Grav., 20, 457

\bibitem[Dadhich \& Turakulov(2002)]{dt02}
Dadhich, N., \& Turakulov, Z. Ya. 2002, Class. Quantum Grav., 19,
2765

\bibitem[Deutsch(1955)]{d55}
Deutsch, A. 1955, Ann. d'Ap., 18, 1

\bibitem[Ginzburg \& Ozernoy(1964)]{go64}
 Ginzburg, V. L., $\&$ Ozernoy, L. M.
1964, Zh. Eksp. Teor. Fiz., 47, 1030

\bibitem[Goldreich \& Julian(1969)]{gj69}
Goldreich, P., \& Julian, W. H. 1969, ApJ, 157, 869

\bibitem[Hartle \& Thorne(1968)]{ht68}
Hartle, J. B., \& Thorne K. S. 1968, ApJ, 153, 807

\bibitem[Kaspi et al.(2006)]{Kaspi04}
Kaspi, V. M., Roberts, M. S. E., \& Harding, A. K. 2006, in
Compact Stellar X-ray Sources, W. Lewin and M. van der Klis, eds.
(Cambridge: Cambridge UP), 279-339 (astro-ph/0402136)

\bibitem[Mestel(1971)]{m71}
Mestel, L. 1971, Nature, 233, 149

\bibitem[Mofiz \& Ahmedov(2000)]{ma00}
Mofiz, U.A., \& Ahmedov, B.J. 2000, ApJ, 542, 484

\bibitem[Muslimov \& Harding(1997)]{MusHar97}
Muslimov, A., $\&$ Harding, A. K. 1997, ApJ, 485, 735

\bibitem[Muslimov \& Tsygan(1986)]{mt86}
Muslimov, A., $\&$ Tsygan, A. I. 1986, AZh, 63, 958

\bibitem[Muslimov \& Tsygan(1990)]{mt90}
Muslimov, A., $\&$ Tsygan, A. I. 1990, Soviet Astr., 34, 133

\bibitem[Muslimov \& Tsygan(1991)a]{mt91}
Muslimov, A., $\&$ Tsygan, A. I. 1991, Preprint No. 1544 of A. F.
Ioffe Institute

\bibitem[Muslimov \& Tsygan(1991)b]{MusTsy91}
Muslimov, A., $\&$ Tsygan, A. I. 1991, in: The Magnetospheric
Structure and Emission Mechanisms of Radio Pulsars, IAU Colloq.
No. 128, p. 340, eds Hankins, T., Rankin, J. \& Gil, J., Kluwer,
Dordrecht

\bibitem[Muslimov \& Tsygan(1992)]{MusTsy92}
Muslimov, A.G., \& Tsygan, A.L. 1992, Mon. Not. R. Astr. Soc.,
255, 61

\bibitem[Newman, Tamburino \& Unti(1963)]{nut63}
Newman, E., Tamburino, L., \& Unti, T. 1963, J. Math. Phys., 4,
915

\bibitem[Rezzolla et al.(2001)]{r1}
Rezzolla, L., Ahmedov, B.J., \& Miller, J.C. 2001a, Mon. Not. R.
    Astron. Soc., 322, 723; Erratum 338, 816 (2003)

\bibitem[Rezzolla et al.(2001)]{r2}
Rezzolla, L., Ahmedov, B.J., \& Miller, J.C. 2001b, Found. of
Phys., 31, 1051

\bibitem[Rezzolla \& Ahmedov(2004)]{r3}
Rezzolla,  L., \& Ahmedov, B.J. 2004, Mon. Not. R. Astron. Soc.,
352, 1161

\bibitem[Ruderman \& Sutherland(1975)]{rs75}
Ruderman, M., $\&$ Sutherland, P. G. 1975, ApJ, 196, 51

\bibitem[Sakai \& Shibata(2003)]{ss03}
Sakai, N., \& Shibata, S. 2003, ApJ, 584, 427

\bibitem[Sturrock(1971)]{s71}
Sturrock, P. A. 1971, ApJ, 164, 179\

\bibitem[Zhu \& Ruderman(1997)]{zhr97}
 Zhu, T.,
$\&$ Ruderman, M. 1997, ApJ, 478, 701

\end{thebibliography}
\end{document}